\documentclass[aps, prl, a4paper, amsfonts, amssymb, amsmath, reprint, showkeys, nofootinbib, twoside,superscriptaddress, ,longbibliography]{revtex4-1}
\usepackage[english]{babel}
\usepackage[utf8]{inputenc}
\usepackage[colorinlistoftodos, color=green!40, prependcaption]{todonotes}
\usepackage{amsthm}
\usepackage{mathtools}
\usepackage{physics}
\usepackage{xcolor}
\usepackage{graphicx}
\usepackage[left=23mm,right=13mm,top=35mm,columnsep=15pt]{geometry} 
\usepackage{adjustbox}
\usepackage{placeins}
\usepackage[T1]{fontenc}
\usepackage{lipsum}
\usepackage{csquotes}
\usepackage[pdftex, pdftitle={Article}, pdfauthor={Author},colorlinks = true,allcolors = blue]{hyperref}
\usepackage{lineno}
\usepackage{mhchem}

\usepackage{overpic}

\begin{document}

\title{Observation of azimuth-dependent suppression of hadron pairs \\in electron scattering off nuclei}

\newcommand*{\ANL}{Argonne National Laboratory, Argonne, Illinois 60439}
\newcommand*{\ANLindex}{1}
\affiliation{\ANL}
\newcommand*{\CSUDH}{California State University, Dominguez Hills, Carson, CA 90747}
\newcommand*{\CSUDHindex}{2}
\affiliation{\CSUDH}
\newcommand*{\CANISIUS}{Canisius College, Buffalo, NY 14208}
\newcommand*{\CANISIUSindex}{3}
\affiliation{\CANISIUS}
\newcommand*{\CMU}{Carnegie Mellon University, Pittsburgh, Pennsylvania 15213}
\newcommand*{\CMUindex}{4}
\affiliation{\CMU}
\newcommand*{\CUA}{Catholic University of America, Washington, D.C. 20064}
\newcommand*{\CUAindex}{5}
\affiliation{\CUA}
\newcommand*{\SACLAY}{IRFU, CEA, Universit\'{e} Paris-Saclay, F-91191 Gif-sur-Yvette, France}
\newcommand*{\SACLAYindex}{6}
\affiliation{\SACLAY}
\newcommand*{\CNU}{Christopher Newport University, Newport News, Virginia 23606}
\newcommand*{\CNUindex}{7}
\affiliation{\CNU}
\newcommand*{\UCONN}{University of Connecticut, Storrs, Connecticut 06269}
\newcommand*{\UCONNindex}{8}
\affiliation{\UCONN}
\newcommand*{\DUKE}{Duke University, Durham, North Carolina 27708-0305}
\newcommand*{\DUKEindex}{9}
\affiliation{\DUKE}
\newcommand*{\DUQUESNE}{Duquesne University, 600 Forbes Avenue, Pittsburgh, PA 15282 }
\newcommand*{\DUQUESNEindex}{10}
\affiliation{\DUQUESNE}
\newcommand*{\FU}{Fairfield University, Fairfield CT 06824}
\newcommand*{\FUindex}{11}
\affiliation{\FU}
\newcommand*{\FERRARAU}{Universit\`{a} di Ferrara, 44121 Ferrara, Italy}
\newcommand*{\FERRARAUindex}{12}
\affiliation{\FERRARAU}
\newcommand*{\FIU}{Florida International University, Miami, Florida 33199}
\newcommand*{\FIUindex}{13}
\affiliation{\FIU}
\newcommand*{\FSU}{Florida State University, Tallahassee, Florida 32306}
\newcommand*{\FSUindex}{14}
\affiliation{\FSU}
\newcommand*{\GWUI}{The George Washington University, Washington, DC 20052}
\newcommand*{\GWUIindex}{15}
\affiliation{\GWUI}
\newcommand*{\INFNFE}{INFN, Sezione di Ferrara, 44100 Ferrara, Italy}
\newcommand*{\INFNFEindex}{16}
\affiliation{\INFNFE}
\newcommand*{\INFNFR}{INFN, Laboratori Nazionali di Frascati, 00044 Frascati, Italy}
\newcommand*{\INFNFRindex}{17}
\affiliation{\INFNFR}
\newcommand*{\INFNGE}{INFN, Sezione di Genova, 16146 Genova, Italy}
\newcommand*{\INFNGEindex}{18}
\affiliation{\INFNGE}
\newcommand*{\INFNRO}{INFN, Sezione di Roma Tor Vergata, 00133 Rome, Italy}
\newcommand*{\INFNROindex}{19}
\affiliation{\INFNRO}
\newcommand*{\INFNTUR}{INFN, Sezione di Torino, 10125 Torino, Italy}
\newcommand*{\INFNTURindex}{20}
\affiliation{\INFNTUR}
\newcommand*{\INFNCAT}{INFN, Sezione di Catania, 95123 Catania, Italy}
\newcommand*{\INFNCATindex}{21}
\affiliation{\INFNCAT}
\newcommand*{\INFNPAV}{INFN, Sezione di Pavia, 27100 Pavia, Italy}
\newcommand*{\INFNPAVindex}{22}
\affiliation{\INFNPAV}
\newcommand*{\ORSAY}{Universit\'{e} Paris-Saclay, CNRS/IN2P3, IJCLab, 91405 Orsay, France}
\newcommand*{\ORSAYindex}{23}
\affiliation{\ORSAY}
\newcommand*{\JMU}{James Madison University, Harrisonburg, Virginia 22807}
\newcommand*{\JMUindex}{24}
\affiliation{\JMU}
\newcommand*{\KNU}{Kyungpook National University, Daegu 41566, Republic of Korea}
\newcommand*{\KNUindex}{25}
\affiliation{\KNU}
\newcommand*{\LAMAR}{Lamar University, 4400 MLK Blvd, PO Box 10046, Beaumont, Texas 77710}
\newcommand*{\LAMARindex}{26}
\affiliation{\LAMAR}
\newcommand*{\MIT}{Massachusetts Institute of Technology, Cambridge, Massachusetts  02139-4307}
\newcommand*{\MITindex}{27}
\affiliation{\MIT}
\newcommand*{\MISS}{Mississippi State University, Mississippi State, MS 39762-5167}
\newcommand*{\MISSindex}{28}
\affiliation{\MISS}
\newcommand*{\ITEP}{National Research Centre Kurchatov Institute - ITEP, Moscow, 117259, Russia}
\newcommand*{\ITEPindex}{29}
\affiliation{\ITEP}
\newcommand*{\UNH}{University of New Hampshire, Durham, New Hampshire 03824-3568}
\newcommand*{\UNHindex}{30}
\affiliation{\UNH}
\newcommand*{\NMSU}{New Mexico State University, PO Box 30001, Las Cruces, NM 88003, USA}
\newcommand*{\NMSUindex}{31}
\affiliation{\NMSU}
\newcommand*{\NSU}{Norfolk State University, Norfolk, Virginia 23504}
\newcommand*{\NSUindex}{32}
\affiliation{\NSU}
\newcommand*{\OHIOU}{Ohio University, Athens, Ohio  45701}
\newcommand*{\OHIOUindex}{33}
\affiliation{\OHIOU}
\newcommand*{\ODU}{Old Dominion University, Norfolk, Virginia 23529}
\newcommand*{\ODUindex}{34}
\affiliation{\ODU}
\newcommand*{\JLUGiessen}{II Physikalisches Institut der Universitaet Giessen, 35392 Giessen, Germany}
\newcommand*{\JLUGiessenindex}{35}
\affiliation{\JLUGiessen}
\newcommand*{\RPI}{Rensselaer Polytechnic Institute, Troy, New York 12180-3590}
\newcommand*{\RPIindex}{36}
\affiliation{\RPI}
\newcommand*{\URICH}{University of Richmond, Richmond, Virginia 23173}
\newcommand*{\URICHindex}{37}
\affiliation{\URICH}
\newcommand*{\ROMAII}{Universit\`{a} di Roma Tor Vergata, 00133 Rome Italy}
\newcommand*{\ROMAIIindex}{38}
\affiliation{\ROMAII}
\newcommand*{\MSU}{Skobeltsyn Institute of Nuclear Physics, Lomonosov Moscow State University, 119234 Moscow, Russia}
\newcommand*{\MSUindex}{39}
\affiliation{\MSU}
\newcommand*{\SCAROLINA}{University of South Carolina, Columbia, South Carolina 29208}
\newcommand*{\SCAROLINAindex}{40}
\affiliation{\SCAROLINA}
\newcommand*{\TEMPLE}{Temple University,  Philadelphia, PA 19122 }
\newcommand*{\TEMPLEindex}{41}
\affiliation{\TEMPLE}
\newcommand*{\JLAB}{Thomas Jefferson National Accelerator Facility, Newport News, Virginia 23606}
\newcommand*{\JLABindex}{42}
\affiliation{\JLAB}
\newcommand*{\UTFSM}{Universidad T\'{e}cnica Federico Santa Mar\'{i}a, Casilla 110-V Valpara\'{i}so, Chile}
\newcommand*{\UTFSMindex}{43}
\affiliation{\UTFSM}
\newcommand*{\BRESCIA}{Universit\`{a} degli Studi di Brescia, 25123 Brescia, Italy}
\newcommand*{\BRESCIAindex}{44}
\affiliation{\BRESCIA}
\newcommand*{\MESSU}{Universit\`{a} degli Studi di Messina, 98166 Messina, Italy}
\newcommand*{\MESSUindex}{45}
\affiliation{\MESSU}
\newcommand*{\UCR}{University of California Riverside, 900 University Avenue, Riverside, CA 92521, USA}
\newcommand*{\UCRindex}{46}
\affiliation{\UCR}
\newcommand*{\GLASGOW}{University of Glasgow, Glasgow G12 8QQ, United Kingdom}
\newcommand*{\GLASGOWindex}{47}
\affiliation{\GLASGOW}
\newcommand*{\YORK}{University of York, York YO10 5DD, United Kingdom}
\newcommand*{\YORKindex}{48}
\affiliation{\YORK}
\newcommand*{\VIRGINIA}{University of Virginia, Charlottesville, Virginia 22901}
\newcommand*{\VIRGINIAindex}{49}
\affiliation{\VIRGINIA}
\newcommand*{\WM}{College of William and Mary, Williamsburg, Virginia 23187-8795}
\newcommand*{\WMindex}{50}
\affiliation{\WM}
\newcommand*{\YEREVAN}{Yerevan Physics Institute, 375036 Yerevan, Armenia}
\newcommand*{\YEREVANindex}{51}
\affiliation{\YEREVAN}

\newcommand*{\NOWISU}{Idaho State University, Pocatello, Idaho 83209}
 %%%%%%%%%%%%%%% END OF Latex Macros for institute addresses  %%%%%%%%%%%%%%%%%%%%%%%%% 

\author {S.J.~Paul} 
\affiliation{\UCR}
\author {S.~Mor\'an}
\affiliation{\UCR}
\author {M.~Arratia}
\affiliation{\UCR}
\affiliation{\JLAB}
\author {A.~El~Alaoui} 
\affiliation{\UTFSM}
\author {H.~Hakobyan} 
\affiliation{\UTFSM}
\author {W.~Brooks} 
\affiliation{\UTFSM}
\author {M.J.~Amaryan} 
\affiliation{\ODU}
\author {W.R. Armstrong} 
\affiliation{\ANL}
\author {H.~Atac} 
\affiliation{\TEMPLE}
\author {L.~Baashen} 
\affiliation{\FIU}
% ...

\collaboration{The CLAS Collaboration}
\noaffiliation

%\date{\today} % Leave empty to omit a date

\begin{abstract}
We present the first measurement of di-hadron angular correlations in electron-nucleus scattering. The data were taken with the CLAS detector and a 5.0 GeV electron beam incident on deuterium, carbon, iron, and lead targets. Relative to deuterium, the nuclear yields of charged-pion pairs show a strong suppression for azimuthally opposite pairs, no suppression for azimuthally nearby pairs, and an enhancement of pairs with large invariant mass. These effects grow with increased nuclear size. The data are qualitatively described by the \textsc{GiBUU} model, which suggests that hadrons form near the nuclear surface and undergo multiple-scattering in nuclei. These results show that angular correlation studies can open a new way to elucidate how hadrons form and interact inside nuclei.

\end{abstract}

\maketitle

\textbf{Introduction.}  The quark-to-hadron transition, called hadronization, remains poorly understood in part due to the great challenge it poses to first-principle calculations in quantum chromodynamics. Studying how hadronization occurs inside large nuclei provides a way to perturb the process to potentially reveal its mechanisms and timescales~\cite{Accardi:2009qv,Accardi:2012qut,Barabanov:2020jvn,Brooks:2020fmf}. It also represents a way to probe the transport properties of particles through nuclei~\cite{Baier:1996sk,Guo:2000nz,Liu:2015obw,Dupre:2015jha,Song:2018szi,Ru:2019qvz,Bai:2020jmd,Guiot:2020vsf}, and tune models needed to interpret neutrino experiments~\cite{Alvarez-Ruso:2017oui,Mosel:2019vhx}.

Scattering experiments with electron beams can help elucidate hadron production by providing control over the energy, $\nu$, and momentum, $\vec q$, transferred in the reaction, which is determined from the scattered electron. Previous studies by HERMES~\cite{Airapetian:2000ks,Airapetian:2003mi,Airapetian:2007vu,Airapetian:2009jy,Airapetian:2011jp} and CLAS~\cite{Daniel:2011nq,Moran:2021} experiments revealed that the production of hadrons is strongly suppressed in nuclei, with a complex dependence on the hadron's energy, transverse momentum, and type. Various aspects of these data agree with different models that include either gluon bremsstrahlung, hadron re-scattering and absorption, or a mixture of these~\cite{Accardi:2009qv}. 

Di-hadron measurements can complement single-hadron studies by providing more kinematic variables and higher sensitivity to nuclear effects such as multiple scattering~\cite{Xing:2012ii,Cougoulic:2017ust,Liang:2008vz,Schafer:2013mza,Boer:2015kxa,Zhang:2019toi,Alrashed:2021csd}. Such variables include angular correlations, which were measured in hadron-collider and fixed-target experiments to probe cold and hot nuclear matter (see Refs.~\cite{Accardi:2009qv,Connors:2017ptx} for reviews). No analogous study has been done with electron beams. 

\begin{figure}[h]
\centering
\includegraphics[width=0.98\columnwidth]{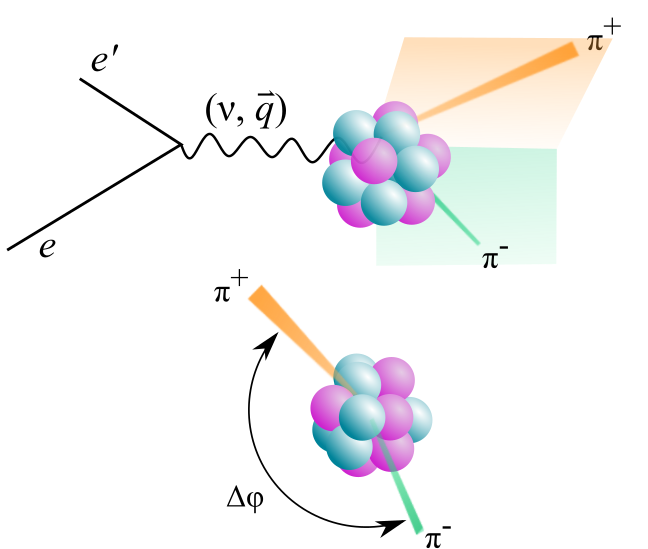}
\caption{Diagram illustrating di-pion production in a nucleus.  Top: side-view, illustrating that the momentum-transfer axis is used to define the azimuthal separation, $\Delta\phi$.  Bottom:  same reaction, as viewed from the direction of the momentum transfer.}
\label{fig:fig1}
\end{figure}

Given the strong absorption of hadrons in nuclei~\cite{Airapetian:2000ks,Airapetian:2003mi,Airapetian:2007vu,Airapetian:2009jy,Airapetian:2011jp,Daniel:2011nq,Moran:2021}, it is expected that most observed hadrons correspond to those that were created near the nuclear surface~\cite{Fialkowski:2007pt} and have their momentum directed away from the center of the nucleus.  Detailed modelling of such geometrical effects remains a challenge~\cite{Liu:2015obw}.  
Di-hadron azimuthal correlations offer a way to test this hypothesis.  When hadron pairs are produced near the surface, the shortest path lengths through the nucleus are obtained when both hadrons' transverse momenta (defined with respect to the momentum-transfer vector, $\vec q$) are directed away from the center of the nucleus relative to their production positions.  When averaging over possible initial-production positions in the nucleus, this favors events where the azimuthal separation between the hadrons, $\Delta\phi$, (also defined with respect to the $\vec q$ direction) is small, while suppressing those with large azimuthal separation
(see Fig.~\ref{fig:fig1}). 
Such a ``surface bias'' has been observed in hadron collisions and has been exploited as a tool to study cold and hot nuclear matter~\cite{Connors:2017ptx}. 

The di-hadron measurements at HERMES~\cite{Airapetian:2005yh} 
revealed hints of nuclear suppression, although with limited precision; moreover, those studies did not explore angular dependence nor did they include hadron identification, which was crucial to elucidate single-hadron studies~\cite{Airapetian:2003mi,Airapetian:2007vu,Airapetian:2011jp}.  Di-hadron measurements were also performed in the SKAT experiment~\cite{Agababyan:2006ne}, which used a neutrino beam incident on heavy nuclei. 

We build upon the HERMES results by studying di-hadron production with a much higher precision and with identified hadrons; in addition, we measure for the first time the di-hadron suppression as a function of the azimuthal separation and the invariant mass. In particular, we study the reaction
\begin{equation}
    eA\rightarrow e'\pi^+\pi^-X,
\end{equation}
where $X$ represents other particles in the event. 

\textbf{Experimental setup.} The data presented here were collected at the Continuous Electron Beam Accelerator Facility (CEBAF) Large Acceptance Spectrometer (CLAS) with a 5.0~GeV electron beam incident on a
 dual-target system~\cite{Hakobyan:2008kua} consisting of a liquid \ce{^{2}H} target
cell and a \ce{^{}C}, \ce{^{}Fe}, or \ce{^{}Pb} foil target.

The CLAS~\cite{Mecking:2003zu} detector was based on a six-fold symmetric toroidal magnet, which defined six sectors instrumented with drift chambers (DC), time-of-flight scintillation counters (TOF), Cherenkov counters (CC), and an electromagnetic calorimeter (EC). Following Refs.~\cite{ElFassi:2012nr,Moran:2021}, electrons were identified by matching negatively charged tracks measured in the DC with hits in the TOF and EC.  Background from $\pi^-$ was suppressed to the $<1\%$ level using the CC and the EC.  Charged pions were identified by checking that the time-of-flight measured from the TOF hits is consistent with the value calculated using the charged-pion mass and the path length and momentum determined with the DCs.
The selection for $\pi^+$ with momentum above 2.7~GeV was further refined by requiring a signal above a certain threshold in the CC to suppress proton background~\cite{Moran:2021}.
Fiducial cuts on momentum and angles were used in order to avoid regions with steeply varying acceptance or low resolution.

\textbf{Event selection and observables.}  The data were selected with a trigger that required at least one electron candidate with momentum $p>500$~MeV. Similar to Ref.~\cite{Moran:2021}, we selected events with $Q^{2}>1$~GeV$^{2}$, $W>2$~GeV and 2.2$<\nu<$4.2~GeV.  Here, $Q^{2}$ is minus the square of the four-momentum transfer, $W=\sqrt{2m_p\nu+m_p^2-Q^{2}}$ where $m_p$ is the mass of a proton, $\nu=E-E'$ is the energy transfer, and $E$ and $E'$ are the beam- and scattered-electron energies. 

Following Ref.~\cite{Airapetian:2005yh}, we selected events with a ``leading'' pion, defined as having fractional energy $z_1=E_h/\nu>0.5$, where $E_h$ is the energy of the pion; we then measured all the other (``secondary'') pions in the event with charge opposite that of the leading pion in the fractional-energy range $0.05<z_2<0.45$.  
In addition, we removed kinematically forbidden events by requiring $|p_1|+|p_2|<\nu$, where $p_1$ and $p_2$ are the momenta of the pion candidates, in order to suppress proton background\footnote{For events with a pion and an ejected proton, the proton's rest energy is present as part of the nucleus in the initial state, so the conservation of energy requires that $E_\pi +KE_p<\nu$, where $KE_p=\sqrt{p_p^2+m_p^2}-m_p$ is the proton's kinetic energy.  This allows ejected protons to have a higher momentum than otherwise possible, therefore allowing such background to be rejected by the $|p_1|+|p_2|<\nu$ cut.}.
The selection included both resonant and non-resonant di-pion production, including exclusive processes, as well as secondary hadrons arising from hadron re-scattering and other nuclear interactions.

We selected particles arising from scattering from either the deuterium or nuclear targets by using the longitudinal vertex position defined by intersecting their trajectories with the beamline. The resulting vertex resolution ensured negligible ambiguity in the target tagging~\cite{Moran:2021}. 

We used the electron, the leading pion, and the sub-leading pion variables to measure the conditional modification factor, $R_{2h}$, defined~\cite{Airapetian:2005yh} as:
\begin{equation}
    R_{2h}(z_2) = \frac{(dN_{2h}^{A}(z_2)/dz_2)/N_{h}^{A}}{(dN_{2h}^{D}(z_2)/dz_2)/N_{h}^{D}}
    \label{eq:R2h}.
\end{equation}
 Here, $(dN_{2h}(z_{2})/dz_2)/N_{h}$ is the ratio of the differential number of selected events with a secondary hadron with energy fraction $z_2$, a leading pion, and an electron, to the total number of selected events with an electron and a leading pion. In other words, $ R_{2h}$ is the nuclear-to-deuterium ratio of the average number of secondary pions per leading pion.
 The superscript indicates that the term is calculated for a nucleus ($A$) or a deuterium ($D$) target. Likewise, we also measured $R_{2h}$ in this work as a function of the azimuthal separation between the pions, $|\Delta\phi|$, and the di-pion invariant mass, 
 \begin{equation}
     m_{\pi\pi} \equiv \sqrt{(P_1+P_2)^2},
 \end{equation}
 where $P_{1}$ and $P_2$ are the four-momenta of the two pions.  

\textbf{Uncertainties.} 
By construction, $R_{2h}$ is a double ratio that benefits from the cancellation of various corrections for detector effects and thus minimizes the associated systematic uncertainties. Moreover, we exploited the dual-target system~\cite{Hakobyan:2008zz}, which by design minimizes systematic uncertainties related to variations in detector response over time by exposing the deuterium and heavier nuclear targets at the same time.  For this reason, when evaluating the A/D ratio, $R_{2h}$ (see Eq.~\ref{eq:R2h}), we used only the deuterium data taken at the same time as the nuclear data used in the numerator.  

We performed studies on various possible sources of systematic uncertainties using data and simulation studies, as described in detail in the Supplementary Material~\cite{supplementary}.  These studies were similar to those used in the single-pion case (see Ref.~\cite{Moran:2021}), and were further checked for the di-pion final state. For the simulation studies we used the \textsc{Pythia 6.319} event generator and the \textsc{GSIM} package~\cite{GSIM}, which is based on \textsc{Geant3}~\cite{Brun:1994aa}, to simulate the response of the CLAS detector and dual-target setup~\cite{Hakobyan:2008kua}. The simulation was tuned to provide a reasonable description of the data. 
The possible sources of systematic uncertainties studied include acceptance effects (2.0\%), event selection (1.4$-$8.3\%), particle mis-identification (0.4$-$3.9\%), and radiative effects (0.3\%).  We found that the acceptance effects due to the different positions of the two targets contribute a systematic uncertainty of about 2\%, which is similar to that of previous studies \cite{Hakobyan:2008zz,Moran:2021}. 
Other sources of systematic uncertainty, such as cross contamination between bins, beam luminosity, trigger efficiency, Coulomb effects, contamination from particles scattering in the walls of the deuterium target,  and time-dependent effects, were found to be negligible.

The systematic uncertainties from different sources were added in quadrature, and totaled to 2.5-3.0\% for most bins. 
However, for some bins, in particular at high $m_{\pi\pi}$, they reached 8.6\%.  In this region, the large systematic uncertainty is largely due to momentum-dependent variations in the efficiency of the CC, which we used to distinguish between high-momentum $\pi^+$ and protons.
For most bins, the systematic uncertainty dominated over the statistical uncertainty, which ranged from 1.1-12\%, with a median value of 2.6\%.

\begin{figure}
    \centering
        
           \begin{overpic}[width=0.98\columnwidth]{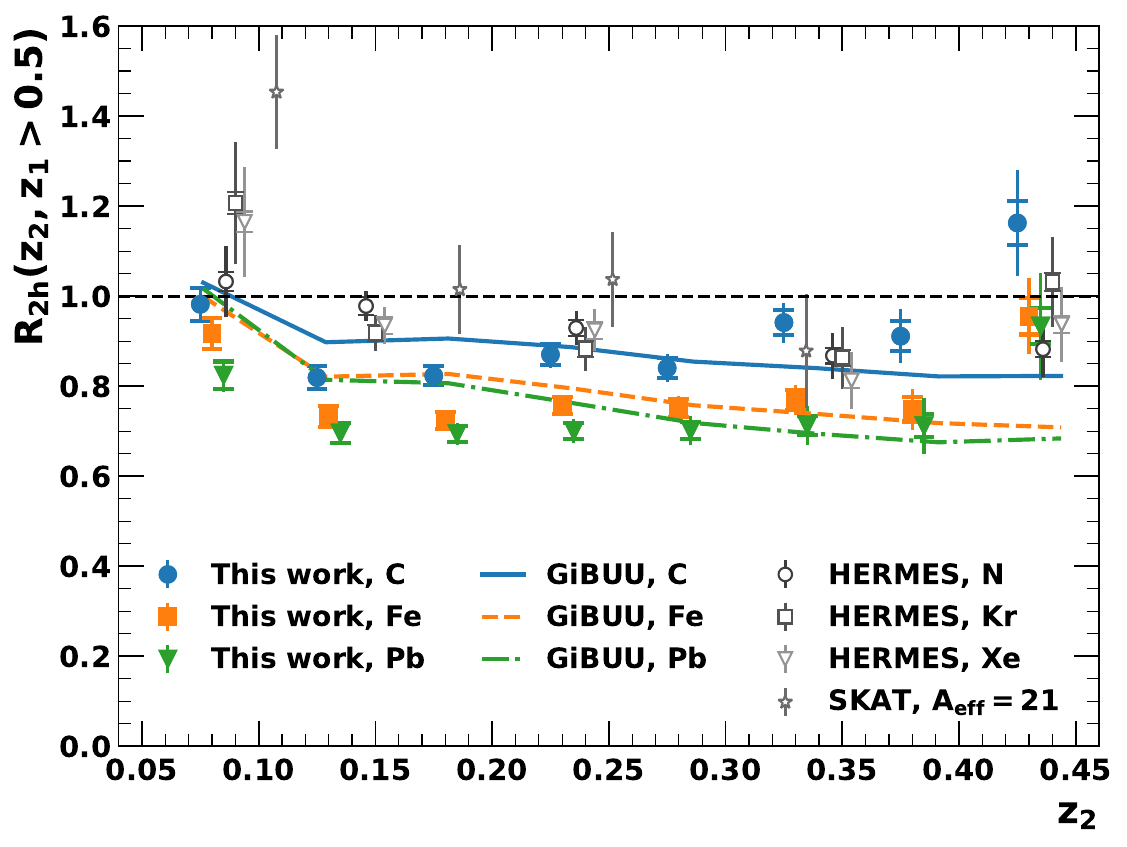}
           \put (17, 65) {\large \textbf{(a)}}
           \end{overpic}
        
        \vspace{0.05cm}
        
            \begin{overpic}[width=0.98\columnwidth]{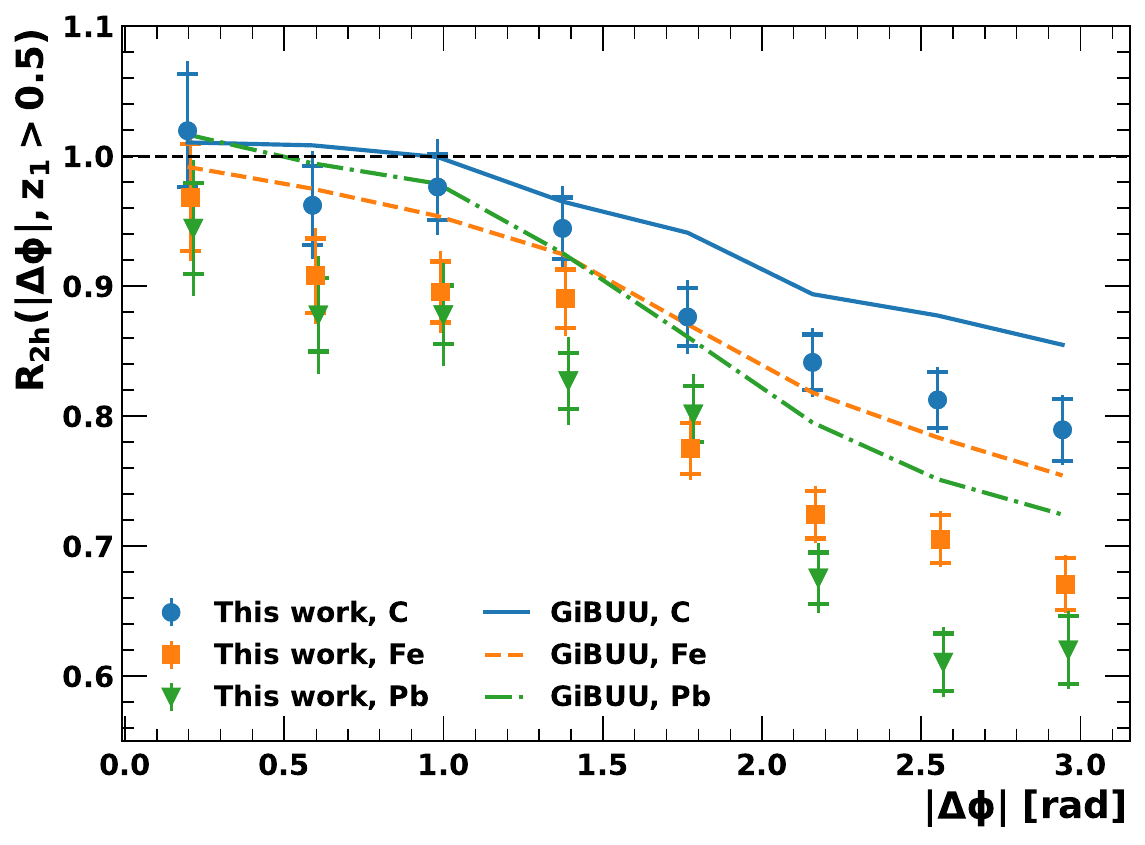}
             \put (17, 65) {\large\textbf{(b)}}
            \end{overpic}
            
        \vspace{0.05cm}
        
                 \begin{overpic}[width=0.98\columnwidth]{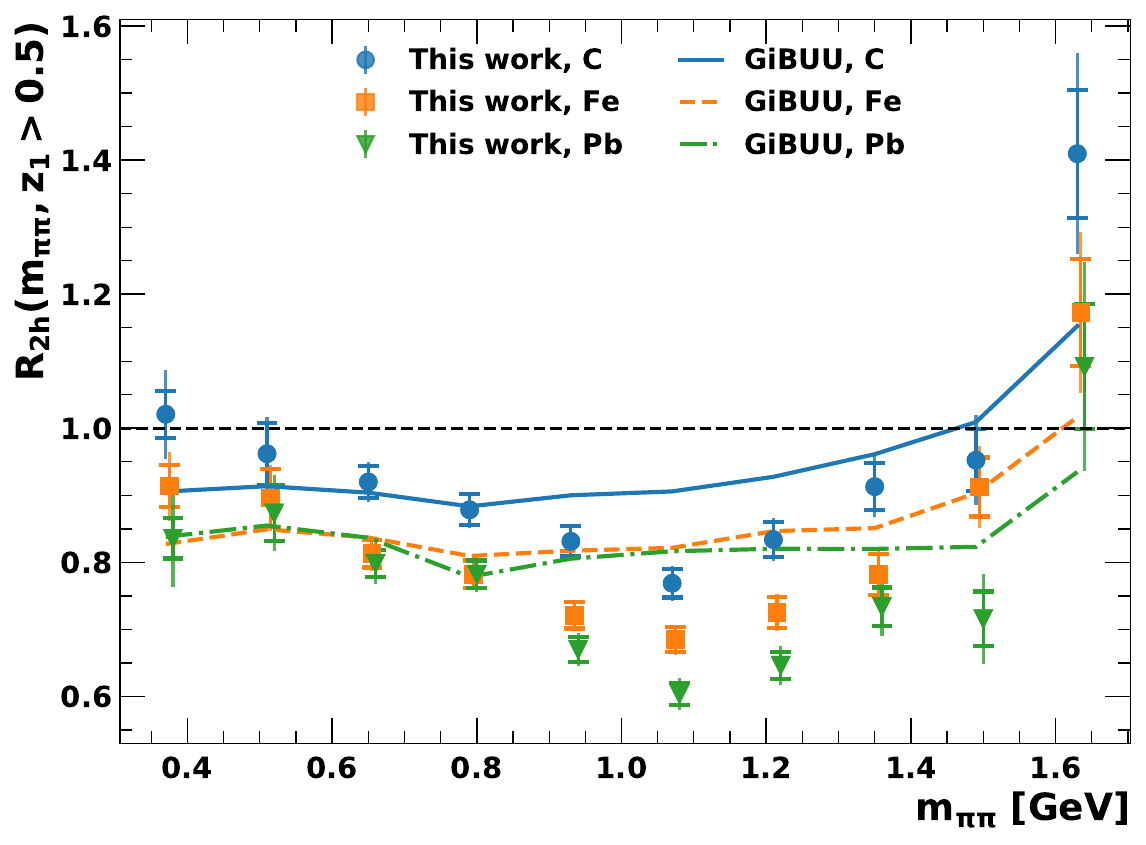}
            \put (17, 65) {\large \textbf{(c)}}
            \end{overpic}
            
              \caption{Conditional suppression factor, $R_{2h}$, as a function of (a) sub-leading hadron $z$, (b) the azimuthal separation $|\Delta\phi|$, and (c) the invariant mass of the pion pair. Points are slighted shifted horizontally for visibility. The gray open symbols in (a) represent results by the HERMES~\cite{Airapetian:2005yh} and SKAT~\cite{Agababyan:2006ne} experiments.  The horizontal caps in the uncertainty bars represent the systematic uncertainties, while the vertical extent of the bars represents the total systematic and statistical uncertainty (added in quadrature). The values of $R_{2h}$, statistical and systematic uncertainties, and bin edges are tabulated in the Supplementary Material~\cite{supplementary}.  Curves represent the calculations from the \textsc{GiBUU} model~\cite{Buss:2011mx}.}
\label{fig:R2withHERMES}
\end{figure}

\textbf{Results and discussion.}  Figure~\ref{fig:R2withHERMES}(a) shows results for $R_{2h}$ as a function of the fractional energy of the sub-leading pion of the pair, $z_{2}$. The CLAS data show a suppression for almost all of the bins, with stronger suppression for heavier nuclei.  $R_{2h}$ depends weakly on $z_2$ except at the first and last bins.  
The values of $R_{2h}$ for Fe and Pb appear close; however, a $\chi^2$ test reveals that their differences are significant at the 99\% CL\footnote{For this test, only the statistical uncertainties were considered, since the systematic uncertainties were assumed to be fully correlated between the different nuclei for any given bin.}.
In the $0.1<z_2<0.4$ range, the average values of $R_{2h}$ are $0.836\pm0.007\pm0.024$, $0.738\pm0.005\pm0.021$, and $0.698\pm0.008\pm0.020$ for \ce{^{}C}, \ce{^{}Fe}, and \ce{^{}Pb}, respectively (where the first uncertainty is statistical and the second is systematic).  We note that the gap between values obtained for C and Fe is much larger than the one between Fe and Pb.  One possible explanation for this is that the event samples with pions in the final state are limited to those that are close to the surface.  If the radius of the nucleus (which scales as $A^{1/3}$) is much larger than the absorption lengths for the pions in nuclei at the relevant kinematics, then this will have a similar effect on both the number of di-pion events, $N^A_{2h}$, and the inclusive number of pion events, $N^A_h$, (see Eq.~\ref{eq:R2h}).  This would then cause the values of the ratio $R_{2h}$ to converge for sufficiently large nuclei.  

Our data in Fig~\ref{fig:R2withHERMES}(a) are compared with existing $eA$ and $\nu A$ data from the HERMES \cite{Airapetian:2005yh} and SKAT \cite{Agababyan:2006ne} experiments, respectively.  
The average kinematics for our results are $\langle\nu\rangle=3.3$~GeV and
$\langle Q^{2} \rangle = 1.6$~GeV$^2$, whereas the HERMES results were at $\langle\nu\rangle = 17.7$~GeV and $\langle Q^{2}\rangle= 2.4$~GeV$^{2}$~\cite{Airapetian:2005yh}. The SKAT data were taken at $\langle\nu\rangle=5.8$~GeV and $\langle Q^{2}\rangle=2.7$~GeV$^{2}$ \cite{Agababyan:2006ne}.  The significant differences observed suggest that the change of kinematics has a strong impact on the nuclear effects. Unlike HERMES, our results show significant evidence for a dependence on the nuclear mass. 

We suggest two effects that could explain the differences between the CLAS and HERMES results.  
First, the smaller energy transfer in the latter experiment causes the hadron-formation length to be shorter.  This increases the distance that the hadrons have to travel to escape, increasing the probability of their absorption.   In the HERMES case, there is an increased probability of pions forming outside of the nucleus, due to longer hadron-formation lengths compared to the CLAS case.  The second explanation is that the pion-nucleon cross sections are larger in the CLAS kinematics (due to lower pion energies than in HERMES).  This would increase the probability of absorption in the CLAS case.  It is also possible that both of these two effects contribute to the differences between the CLAS and HERMES results.  

We compare our data with with the \textsc{GiBUU} Monte-Carlo event generator~\cite{Buss:2011mx} (using the 2019 default parameters), which incorporates treatment of final-state interactions, absorption, and production mechanisms with elastic and inelastic channels. The \textsc{GiBUU} model described reasonably well the single-hadron data from CLAS~\cite{Moran:2021} and HERMES~\cite{Airapetian:2003mi,Airapetian:2007vu,Airapetian:2011jp}. While the \textsc{GiBUU} calculations reproduce some of the qualitative features of the data in this work, including the larger gap between the $R_{2h}$ values for C and Fe than the one between Fe and Pb, there are  significant differences, for instance it predicts an uptick only at low $z_2$, while the data have upticks at both the lowest and the highest $z_2$ bins.  
The low-$z_2$ uptick was also observed in the HERMES and SKAT data, which were at very different kinematics, suggesting that this effect does not depend strongly on $Q^{2}$ or $\nu$.
In the \textsc{GiBUU} model, the uptick in $R_{2h}$ at low $z_2$ is caused by the interaction between hadrons produced in the primary electron-nucleon interaction with other hadrons as they propagate through nuclei. The uptick at high $z_{2}$ is consistent with unity and also exists in the HERMES results.   This high-$z_2$ uptick, which is not reproduced by the \textsc{GiBUU} model, may be due to coherent production in the  $z_1+z_2\rightarrow1$ limit; coherent production in general is not included in the \textsc{GiBUU} model \cite{Buss:2011mx}.

Figure~\ref{fig:R2withHERMES}(b) shows $R_{2h}$ as a function of the azimuthal separation, $\Delta\phi$, between the two pions, as measured around the direction of the momentum transfer (see Fig.~\ref{fig:fig1}). 
The data show significant dependence on $\Delta\phi$ for all nuclei. 
For all three nuclei, the deviation of $R_{2h}$ from unity is smallest when $\Delta\phi$ is near 0 and drops off with increasing $|\Delta\phi|$, with a steeper slope for heavier nuclei.
For azimuthally opposite pairs ($|\Delta\phi|$ near $\pi$), $R_{2h}$ is $0.789\pm0.013\pm0.024$, $0.671\pm0.010\pm0.020$, and $0.620\pm0.015\pm0.026$ for \ce{^{}C}, \ce{^{}Fe}, and \ce{^{}Pb}, respectively. This is qualitatively described by the \textsc{GiBUU} model; however, the data show a more pronounced $\Delta\phi$ dependence.  

We also present $R_{2h}$ as a function of the di-pion invariant mass, $m_{\pi\pi}$, in Fig.~\ref{fig:R2withHERMES}(c). The data show a negative slope in the $0.4<m_{\pi\pi}<1.1$~GeV region, and an enhancement at higher $m_{\pi\pi}$.  We also observe that within the region of negative slope, the dependence appears to be smooth and no abrupt behavior is observed in $R_{2h}$ near the $\rho^0(770)$ mass.  
The data are qualitatively described by \textsc{GiBUU}, including the uptick at high mass. In the \textsc{GiBUU} model, this is caused by re-scattering of hadrons, which leads to larger transverse momentum of hadrons and higher pair invariant mass. The qualitative behavior of the data is reminiscent of the enhancement of hadrons with high-transverse momentum reported in Refs.~\cite{HERMES:2009uge,Airapetian:2000ks,Airapetian:2003mi,Airapetian:2007vu,Airapetian:2009jy,Airapetian:2011jp,Moran:2021}.

Hadron-absorption effects in the \textsc{GiBUU} model can be investigated through looking at the distribution of the hadron-production points of the final-state pions.  We observe that in the \textsc{GiBUU} simulations, a larger fraction of the total final-state pions are formed near the surface of the nucleus than would be expected if their production points were distributed uniformly.  Further, we observe that this effect is stronger for larger nuclei than for smaller ones.  The effect is also stronger in the di-pion case than for the single-pion case.   The latter effect can be explained by the fact that we require the survival of not only the secondary pion but also the leading one as well, biasing the sample further towards the surface.  Such an interpretation is consistent with Ref.~\cite{Fialkowski:2007pt}, which argued that an absorption model coupled with geometrical biases caused by survivor selection could explain both the single and double-hadron data from HERMES without resorting to other effects such as gluon bremsstrahlung~\cite{Majumder:2008jy,Majumder:2009ge}. 

\textbf{Summary and conclusions.} In summary, we have presented a measurement of di-pion production in electron scattering off nuclei using the CLAS detector, which includes the first study on the azimuthal separation and invariant mass. 
The data show a strong suppression for azimuthally opposite pairs, no suppression for pairs with small azimuthal separation, and an enhancement of pairs with large invariant mass. This is qualitatively consistent with the predictions from the \textsc{GiBUU} model, wherein it can be attributed to an increased probability of absorption of hadrons in azimuthally opposite pairs due to the increased path length compared to azimuthally nearby pairs; however, the measured suppression is stronger than in the predictions.  

We also measured the dependence of the nuclear-to-deuterium ratio on the secondary pion's fractional energy, and on the di-pion invariant mass.  We compared our measurement of the dependence on the former with results from HERMES and found both to be qualitatively comparable; however, our measurement shows a stronger nuclear dependence than the HERMES results, suggesting a strong kinematic dependence of the observed effects.  Further, we note that the di-hadron suppression is stronger in heavier nuclei, although the effect appears to saturate for higher nuclear masses.  

Overall, the data show evidence that nuclear effects not only modify the hadron distributions, but also modify the correlations between multiple hadrons in the event, relative to the correlations that exist in the deuteron case due to momentum conservation and limited phase space. 

Our studies show how kinematic variables that depend on both hadrons, such as azimuthal separation and pair mass, can be used as a powerful tool for studying hadron production in electron scattering off nuclei. Given that these data cover a poorly explored kinematic domain where the hadron-formation length is expected to be similar to the nuclear size, future comparisons to models (similar to those of Refs. \cite{GALLMEISTER200868,PhysRevC.101.014617}) might shed light on hadronization timescales and mechanisms.

These results provide a reference for planned di-hadron measurements in future experiments with higher beam energies at the Jefferson Laboratory~\cite{Burkert:2018nvj,Burkert:2020akg,Arrington:2021alx}, and future electron-ion colliders in the USA~\cite{Accardi:2012qut,AbdulKhalek:2021gbh} and China~\cite{Anderle:2021wcy}.

\begin{acknowledgements}
\textbf{Acknowledgements.} The authors acknowledge the staff of the Accelerator and Physics Divisions at 
the Thomas Jefferson National Accelerator Facility who made this experiment 
possible.
We thank Kai Gallmeister for help in setting up the \textsc{GiBUU} event generator. This work was supported in part by the Chilean Agencia Nacional de Investigacion y Desarollo (ANID), by ANID PIA grant 
ACT1413, by ANID PIA/APOYO AFB180002, by ANID FONDECYT No. 1161642 and No. 1201964 
and No. 11181215 and No. 1151248, by the ANID-Millennium Science Initiative Program - ICN2019\_044, by the U.S.
Department of Energy, the Italian Instituto Nazionale di Fisica Nucleare, the French Centre 
National de la Recherche Scientifique, the French Commissariat \`a l'Energie 
Atomique, 
the United Kingdom Science and Technology Facilities Council (STFC), the 
Scottish Universities Physics Alliance (SUPA), the National Research Foundation 
of Korea, the National Science Foundation (NSF),  the HelmholtzForschungsakademie Hessen für FAIR (HFHF), the Ministry of Science and Higher Education of the Russian Federation, and the Office of Research and Economic Development at Mississippi 
State University. This work has received funding from 
the European Research Council (ERC) under the European Union’s Horizon 2020 
research and innovation programme (Grant agreement No. 804480). The Southeastern Universities Research 
Association operates the Thomas Jefferson National Accelerator Facility for the 
United States Department of Energy under Contract No. DE-AC05-06OR23177.
\end{acknowledgements}

 \FloatBarrier
\bibliographystyle{apsrev4-1}
\bibliography{biblio.bib} % refers to example.bib

\begin{thebibliography}{50}%
\makeatletter
\providecommand \@ifxundefined [1]{%
 \@ifx{#1\undefined}
}%
\providecommand \@ifnum [1]{%
 \ifnum #1\expandafter \@firstoftwo
 \else \expandafter \@secondoftwo
 \fi
}%
\providecommand \@ifx [1]{%
 \ifx #1\expandafter \@firstoftwo
 \else \expandafter \@secondoftwo
 \fi
}%
\providecommand \natexlab [1]{#1}%
\providecommand \enquote  [1]{``#1''}%
\providecommand \bibnamefont  [1]{#1}%
\providecommand \bibfnamefont [1]{#1}%
\providecommand \citenamefont [1]{#1}%
\providecommand \href@noop [0]{\@secondoftwo}%
\providecommand \href [0]{\begingroup \@sanitize@url \@href}%
\providecommand \@href[1]{\@@startlink{#1}\@@href}%
\providecommand \@@href[1]{\endgroup#1\@@endlink}%
\providecommand \@sanitize@url [0]{\catcode `\\12\catcode `\$12\catcode
  `\&12\catcode `\#12\catcode `\^12\catcode `\_12\catcode `\%12\relax}%
\providecommand \@@startlink[1]{}%
\providecommand \@@endlink[0]{}%
\providecommand \url  [0]{\begingroup\@sanitize@url \@url }%
\providecommand \@url [1]{\endgroup\@href {#1}{\urlprefix }}%
\providecommand \urlprefix  [0]{URL }%
\providecommand \Eprint [0]{\href }%
\providecommand \doibase [0]{http://dx.doi.org/}%
\providecommand \selectlanguage [0]{\@gobble}%
\providecommand \bibinfo  [0]{\@secondoftwo}%
\providecommand \bibfield  [0]{\@secondoftwo}%
\providecommand \translation [1]{[#1]}%
\providecommand \BibitemOpen [0]{}%
\providecommand \bibitemStop [0]{}%
\providecommand \bibitemNoStop [0]{.\EOS\space}%
\providecommand \EOS [0]{\spacefactor3000\relax}%
\providecommand \BibitemShut  [1]{\csname bibitem#1\endcsname}%
\let\auto@bib@innerbib\@empty
%</preamble>
\bibitem [{\citenamefont {Accardi}\ \emph {et~al.}(2009)\citenamefont
  {Accardi}, \citenamefont {Arleo}, \citenamefont {Brooks}, \citenamefont
  {D'Enterria},\ and\ \citenamefont {Muccifora}}]{Accardi:2009qv}%
  \BibitemOpen
  \bibfield  {author} {\bibinfo {author} {\bibfnamefont {A.}~\bibnamefont
  {Accardi}}, \bibinfo {author} {\bibfnamefont {F.}~\bibnamefont {Arleo}},
  \bibinfo {author} {\bibfnamefont {W.~K.}\ \bibnamefont {Brooks}}, \bibinfo
  {author} {\bibfnamefont {D.}~\bibnamefont {D'Enterria}}, \ and\ \bibinfo
  {author} {\bibfnamefont {V.}~\bibnamefont {Muccifora}},\ }\href {\doibase
  10.1393/ncr/i2009-10048-0} {\bibfield  {journal} {\bibinfo  {journal} {Riv.
  Nuovo Cim.}\ }\textbf {\bibinfo {volume} {32}},\ \bibinfo {pages} {439}
  (\bibinfo {year} {2009})}\BibitemShut {NoStop}%
%%CITATION = ARXIV:0907.3534;%%
\bibitem [{\citenamefont {Accardi}\ \emph {et~al.}(2016)\citenamefont {Accardi}
  \emph {et~al.}}]{Accardi:2012qut}%
  \BibitemOpen
  \bibfield  {author} {\bibinfo {author} {\bibfnamefont {A.}~\bibnamefont
  {Accardi}} \emph {et~al.},\ }\href {\doibase 10.1140/epja/i2016-16268-9}
  {\bibfield  {journal} {\bibinfo  {journal} {Eur. Phys. J.}\ }\textbf
  {\bibinfo {volume} {A52}},\ \bibinfo {pages} {268} (\bibinfo {year}
  {2016})}\BibitemShut {NoStop}%
%%CITATION = ARXIV:1212.1701;%%
\bibitem [{\citenamefont {Barabanov}\ \emph {et~al.}(2021)\citenamefont
  {Barabanov} \emph {et~al.}}]{Barabanov:2020jvn}%
  \BibitemOpen
  \bibfield  {author} {\bibinfo {author} {\bibfnamefont {M.~Y.}\ \bibnamefont
  {Barabanov}} \emph {et~al.},\ }\href {\doibase 10.1016/j.ppnp.2020.103835}
  {\bibfield  {journal} {\bibinfo  {journal} {Prog. Part. Nucl. Phys.}\
  }\textbf {\bibinfo {volume} {116}},\ \bibinfo {pages} {103835} (\bibinfo
  {year} {2021})}\BibitemShut {NoStop}%
\bibitem [{\citenamefont {Brooks}\ and\ \citenamefont
  {L\'opez}(2021)}]{Brooks:2020fmf}%
  \BibitemOpen
  \bibfield  {author} {\bibinfo {author} {\bibfnamefont {W.~K.}\ \bibnamefont
  {Brooks}}\ and\ \bibinfo {author} {\bibfnamefont {J.~A.}\ \bibnamefont
  {L\'opez}},\ }\href {\doibase 10.1016/j.physletb.2021.136171} {\bibfield
  {journal} {\bibinfo  {journal} {Phys. Lett. B}\ }\textbf {\bibinfo {volume}
  {816}},\ \bibinfo {pages} {136171} (\bibinfo {year} {2021})}\BibitemShut
  {NoStop}%
\bibitem [{\citenamefont {Baier}\ \emph {et~al.}(1997)\citenamefont {Baier},
  \citenamefont {Dokshitzer}, \citenamefont {Mueller}, \citenamefont
  {Peign\'e},\ and\ \citenamefont {Schiff}}]{Baier:1996sk}%
  \BibitemOpen
  \bibfield  {author} {\bibinfo {author} {\bibfnamefont {R.}~\bibnamefont
  {Baier}}, \bibinfo {author} {\bibfnamefont {Y.~L.}\ \bibnamefont
  {Dokshitzer}}, \bibinfo {author} {\bibfnamefont {A.~H.}\ \bibnamefont
  {Mueller}}, \bibinfo {author} {\bibfnamefont {S.}~\bibnamefont {Peign\'e}}, \
  and\ \bibinfo {author} {\bibfnamefont {D.}~\bibnamefont {Schiff}},\ }\href
  {\doibase 10.1016/S0550-3213(96)00581-0} {\bibfield  {journal} {\bibinfo
  {journal} {Nucl. Phys. B}\ }\textbf {\bibinfo {volume} {484}},\ \bibinfo
  {pages} {265} (\bibinfo {year} {1997})}\BibitemShut {NoStop}%
\bibitem [{\citenamefont {Guo}\ and\ \citenamefont {Wang}(2000)}]{Guo:2000nz}%
  \BibitemOpen
  \bibfield  {author} {\bibinfo {author} {\bibfnamefont {X.}~\bibnamefont
  {Guo}}\ and\ \bibinfo {author} {\bibfnamefont {X.-N.}\ \bibnamefont {Wang}},\
  }\href {\doibase 10.1103/PhysRevLett.85.3591} {\bibfield  {journal} {\bibinfo
   {journal} {Phys. Rev. Lett.}\ }\textbf {\bibinfo {volume} {85}},\ \bibinfo
  {pages} {3591} (\bibinfo {year} {2000})}\BibitemShut {NoStop}%
\bibitem [{\citenamefont {Liu}\ \emph {et~al.}(2015)\citenamefont {Liu},
  \citenamefont {Miao}, \citenamefont {Song},\ and\ \citenamefont
  {Duan}}]{Liu:2015obw}%
  \BibitemOpen
  \bibfield  {author} {\bibinfo {author} {\bibfnamefont {N.}~\bibnamefont
  {Liu}}, \bibinfo {author} {\bibfnamefont {W.-D.}\ \bibnamefont {Miao}},
  \bibinfo {author} {\bibfnamefont {L.-H.}\ \bibnamefont {Song}}, \ and\
  \bibinfo {author} {\bibfnamefont {C.-G.}\ \bibnamefont {Duan}},\ }\href
  {\doibase 10.1016/j.physletb.2015.07.048} {\bibfield  {journal} {\bibinfo
  {journal} {Phys. Lett.}\ }\textbf {\bibinfo {volume} {B749}},\ \bibinfo
  {pages} {88} (\bibinfo {year} {2015})}\BibitemShut {NoStop}%
%%CITATION = ARXIV:1511.00767;%%
\bibitem [{\citenamefont {Dupr\'e}\ and\ \citenamefont
  {Scopetta}(2016)}]{Dupre:2015jha}%
  \BibitemOpen
  \bibfield  {author} {\bibinfo {author} {\bibfnamefont {R.}~\bibnamefont
  {Dupr\'e}}\ and\ \bibinfo {author} {\bibfnamefont {S.}~\bibnamefont
  {Scopetta}},\ }\href {\doibase 10.1140/epja/i2016-16159-1} {\bibfield
  {journal} {\bibinfo  {journal} {Eur. Phys. J. A}\ }\textbf {\bibinfo {volume}
  {52}},\ \bibinfo {pages} {159} (\bibinfo {year} {2016})},\ \Eprint
  {http://arxiv.org/abs/1510.00794} {arXiv:1510.00794 [nucl-ex]} \BibitemShut
  {NoStop}%
\bibitem [{\citenamefont {Song}\ \emph {et~al.}(2018)\citenamefont {Song},
  \citenamefont {Xin},\ and\ \citenamefont {Liu}}]{Song:2018szi}%
  \BibitemOpen
  \bibfield  {author} {\bibinfo {author} {\bibfnamefont {L.-H.}\ \bibnamefont
  {Song}}, \bibinfo {author} {\bibfnamefont {S.-F.}\ \bibnamefont {Xin}}, \
  and\ \bibinfo {author} {\bibfnamefont {N.}~\bibnamefont {Liu}},\ }\href
  {\doibase 10.1088/1361-6471/aaa09b} {\bibfield  {journal} {\bibinfo
  {journal} {J. Phys.}\ }\textbf {\bibinfo {volume} {G45}},\ \bibinfo {pages}
  {025005} (\bibinfo {year} {2018})}\BibitemShut {NoStop}%
%%CITATION = JPAGA,G45,025005;%%
\bibitem [{\citenamefont {Ru}\ \emph {et~al.}(2021)\citenamefont {Ru},
  \citenamefont {Kang}, \citenamefont {Wang}, \citenamefont {Xing},\ and\
  \citenamefont {Zhang}}]{Ru:2019qvz}%
  \BibitemOpen
  \bibfield  {author} {\bibinfo {author} {\bibfnamefont {P.}~\bibnamefont
  {Ru}}, \bibinfo {author} {\bibfnamefont {Z.-B.}\ \bibnamefont {Kang}},
  \bibinfo {author} {\bibfnamefont {E.}~\bibnamefont {Wang}}, \bibinfo {author}
  {\bibfnamefont {H.}~\bibnamefont {Xing}}, \ and\ \bibinfo {author}
  {\bibfnamefont {B.-W.}\ \bibnamefont {Zhang}},\ }\href {\doibase
  10.1103/PhysRevD.103.L031901} {\bibfield  {journal} {\bibinfo  {journal}
  {Phys. Rev. D}\ }\textbf {\bibinfo {volume} {103}},\ \bibinfo {pages}
  {L031901} (\bibinfo {year} {2021})}\BibitemShut {NoStop}%
\bibitem [{\citenamefont {Bai}\ and\ \citenamefont {Duan}(2021)}]{Bai:2020jmd}%
  \BibitemOpen
  \bibfield  {author} {\bibinfo {author} {\bibfnamefont {T.-X.}\ \bibnamefont
  {Bai}}\ and\ \bibinfo {author} {\bibfnamefont {C.-G.}\ \bibnamefont {Duan}},\
  }\href {\doibase 10.1140/epjp/s13360-021-02174-5} {\bibfield  {journal}
  {\bibinfo  {journal} {Eur. Phys. J. Plus}\ }\textbf {\bibinfo {volume}
  {136}},\ \bibinfo {pages} {1181} (\bibinfo {year} {2021})}\BibitemShut
  {NoStop}%
\bibitem [{\citenamefont {Guiot}\ and\ \citenamefont
  {Kopeliovich}(2020)}]{Guiot:2020vsf}%
  \BibitemOpen
  \bibfield  {author} {\bibinfo {author} {\bibfnamefont {B.}~\bibnamefont
  {Guiot}}\ and\ \bibinfo {author} {\bibfnamefont {B.~Z.}\ \bibnamefont
  {Kopeliovich}},\ }\href {\doibase 10.1103/PhysRevC.102.045201} {\bibfield
  {journal} {\bibinfo  {journal} {Phys. Rev. C}\ }\textbf {\bibinfo {volume}
  {102}},\ \bibinfo {pages} {045201} (\bibinfo {year} {2020})}\BibitemShut
  {NoStop}%
\bibitem [{\citenamefont {Alvarez-Ruso}\ \emph {et~al.}(2018)\citenamefont
  {Alvarez-Ruso} \emph {et~al.}}]{Alvarez-Ruso:2017oui}%
  \BibitemOpen
  \bibfield  {author} {\bibinfo {author} {\bibfnamefont {L.}~\bibnamefont
  {Alvarez-Ruso}} \emph {et~al.},\ }\href {\doibase 10.1016/j.ppnp.2018.01.006}
  {\bibfield  {journal} {\bibinfo  {journal} {Prog. Part. Nucl. Phys.}\
  }\textbf {\bibinfo {volume} {100}},\ \bibinfo {pages} {1} (\bibinfo {year}
  {2018})}\BibitemShut {NoStop}%
%%CITATION = ARXIV:1706.03621;%%
\bibitem [{\citenamefont {Mosel}(2019)}]{Mosel:2019vhx}%
  \BibitemOpen
  \bibfield  {author} {\bibinfo {author} {\bibfnamefont {U.}~\bibnamefont
  {Mosel}},\ }\href {\doibase 10.1088/1361-6471/ab3830} {\bibfield  {journal}
  {\bibinfo  {journal} {J. Phys.}\ }\textbf {\bibinfo {volume} {G46}},\
  \bibinfo {pages} {113001} (\bibinfo {year} {2019})}\BibitemShut {NoStop}%
%%CITATION = ARXIV:1904.11506;%%
\bibitem [{\citenamefont {Airapetian}\ \emph {et~al.}(2001)\citenamefont
  {Airapetian} \emph {et~al.}}]{Airapetian:2000ks}%
  \BibitemOpen
  \bibfield  {author} {\bibinfo {author} {\bibfnamefont {A.}~\bibnamefont
  {Airapetian}} \emph {et~al.} (\bibinfo {collaboration} {\textit{HERMES
  Collaboration}}),\ }\href {\doibase 10.1007/s100520100697} {\bibfield
  {journal} {\bibinfo  {journal} {Eur. Phys. J.}\ }\textbf {\bibinfo {volume}
  {C20}},\ \bibinfo {pages} {479} (\bibinfo {year} {2001})}\BibitemShut
  {NoStop}%
%%CITATION = HEP-EX/0012049;%%
\bibitem [{\citenamefont {Airapetian}\ \emph {et~al.}(2003)\citenamefont
  {Airapetian} \emph {et~al.}}]{Airapetian:2003mi}%
  \BibitemOpen
  \bibfield  {author} {\bibinfo {author} {\bibfnamefont {A.}~\bibnamefont
  {Airapetian}} \emph {et~al.} (\bibinfo {collaboration} {\textit{HERMES
  Collaboration}}),\ }\href {\doibase 10.1016/j.physletb.2003.10.026}
  {\bibfield  {journal} {\bibinfo  {journal} {Phys. Lett.}\ }\textbf {\bibinfo
  {volume} {B577}},\ \bibinfo {pages} {37} (\bibinfo {year}
  {2003})}\BibitemShut {NoStop}%
%%CITATION = HEP-EX/0307023;%%
\bibitem [{\citenamefont {Airapetian}\ \emph {et~al.}(2007)\citenamefont
  {Airapetian} \emph {et~al.}}]{Airapetian:2007vu}%
  \BibitemOpen
  \bibfield  {author} {\bibinfo {author} {\bibfnamefont {A.}~\bibnamefont
  {Airapetian}} \emph {et~al.} (\bibinfo {collaboration} {\textit{HERMES
  Collaboration}}),\ }\href {\doibase 10.1016/j.nuclphysb.2007.06.004}
  {\bibfield  {journal} {\bibinfo  {journal} {Nucl. Phys.}\ }\textbf {\bibinfo
  {volume} {B780}},\ \bibinfo {pages} {1} (\bibinfo {year} {2007})}\BibitemShut
  {NoStop}%
%%CITATION = ARXIV:0704.3270;%%
\bibitem [{\citenamefont {Airapetian}\ \emph
  {et~al.}(2010{\natexlab{a}})\citenamefont {Airapetian} \emph
  {et~al.}}]{Airapetian:2009jy}%
  \BibitemOpen
  \bibfield  {author} {\bibinfo {author} {\bibfnamefont {A.}~\bibnamefont
  {Airapetian}} \emph {et~al.} (\bibinfo {collaboration} {\textit{HERMES
  Collaboration}}),\ }\href {\doibase 10.1016/j.physletb.2010.01.020}
  {\bibfield  {journal} {\bibinfo  {journal} {Phys. Lett.}\ }\textbf {\bibinfo
  {volume} {B684}},\ \bibinfo {pages} {114} (\bibinfo {year}
  {2010}{\natexlab{a}})}\BibitemShut {NoStop}%
%%CITATION = ARXIV:0906.2478;%%
\bibitem [{\citenamefont {Airapetian}\ \emph {et~al.}(2011)\citenamefont
  {Airapetian} \emph {et~al.}}]{Airapetian:2011jp}%
  \BibitemOpen
  \bibfield  {author} {\bibinfo {author} {\bibfnamefont {A.}~\bibnamefont
  {Airapetian}} \emph {et~al.} (\bibinfo {collaboration} {\textit{HERMES
  Collaboration}}),\ }\href {\doibase 10.1140/epja/i2011-11113-5} {\bibfield
  {journal} {\bibinfo  {journal} {Eur. Phys. J.}\ }\textbf {\bibinfo {volume}
  {A47}},\ \bibinfo {pages} {113} (\bibinfo {year} {2011})}\BibitemShut
  {NoStop}%
%%CITATION = ARXIV:1107.3496;%%
\bibitem [{\citenamefont {Daniel}\ \emph {et~al.}(2011)\citenamefont {Daniel}
  \emph {et~al.}}]{Daniel:2011nq}%
  \BibitemOpen
  \bibfield  {author} {\bibinfo {author} {\bibfnamefont {A.}~\bibnamefont
  {Daniel}} \emph {et~al.} (\bibinfo {collaboration} {\textit{CLAS
  Collaboration}}),\ }\href {\doibase 10.1016/j.physletb.2011.10.071}
  {\bibfield  {journal} {\bibinfo  {journal} {Phys. Lett.}\ }\textbf {\bibinfo
  {volume} {B706}},\ \bibinfo {pages} {26} (\bibinfo {year}
  {2011})}\BibitemShut {NoStop}%
%%CITATION = ARXIV:1111.2573;%%
\bibitem [{\citenamefont {Mor\'an}\ \emph {et~al.}(2022)\citenamefont {Mor\'an}
  \emph {et~al.}}]{Moran:2021}%
  \BibitemOpen
  \bibfield  {author} {\bibinfo {author} {\bibfnamefont {S.}~\bibnamefont
  {Mor\'an}} \emph {et~al.} (\bibinfo {collaboration} {\textit{CLAS
  Collaboration}}),\ }\href {\doibase 10.1103/PhysRevC.105.015201} {\bibfield
  {journal} {\bibinfo  {journal} {Phys. Rev. C}\ }\textbf {\bibinfo {volume}
  {105}},\ \bibinfo {pages} {015201} (\bibinfo {year} {2022})}\BibitemShut
  {NoStop}%
\bibitem [{\citenamefont {Xing}\ \emph {et~al.}(2012)\citenamefont {Xing},
  \citenamefont {Kang}, \citenamefont {Vitev},\ and\ \citenamefont
  {Wang}}]{Xing:2012ii}%
  \BibitemOpen
  \bibfield  {author} {\bibinfo {author} {\bibfnamefont {H.}~\bibnamefont
  {Xing}}, \bibinfo {author} {\bibfnamefont {Z.-B.}\ \bibnamefont {Kang}},
  \bibinfo {author} {\bibfnamefont {I.}~\bibnamefont {Vitev}}, \ and\ \bibinfo
  {author} {\bibfnamefont {E.}~\bibnamefont {Wang}},\ }\href {\doibase
  10.1103/PhysRevD.86.094010} {\bibfield  {journal} {\bibinfo  {journal} {Phys.
  Rev. D}\ }\textbf {\bibinfo {volume} {86}},\ \bibinfo {pages} {094010}
  (\bibinfo {year} {2012})}\BibitemShut {NoStop}%
\bibitem [{\citenamefont {Cougoulic}\ and\ \citenamefont
  {Peign\'e}(2018)}]{Cougoulic:2017ust}%
  \BibitemOpen
  \bibfield  {author} {\bibinfo {author} {\bibfnamefont {F.}~\bibnamefont
  {Cougoulic}}\ and\ \bibinfo {author} {\bibfnamefont {S.}~\bibnamefont
  {Peign\'e}},\ }\href {\doibase 10.1007/JHEP05(2018)203} {\bibfield  {journal}
  {\bibinfo  {journal} {JHEP}\ }\textbf {\bibinfo {volume} {05}},\ \bibinfo
  {pages} {203} (\bibinfo {year} {2018})}\BibitemShut {NoStop}%
\bibitem [{\citenamefont {Liang}\ \emph {et~al.}(2008)\citenamefont {Liang},
  \citenamefont {Wang},\ and\ \citenamefont {Zhou}}]{Liang:2008vz}%
  \BibitemOpen
  \bibfield  {author} {\bibinfo {author} {\bibfnamefont {Z.-t.}\ \bibnamefont
  {Liang}}, \bibinfo {author} {\bibfnamefont {X.-N.}\ \bibnamefont {Wang}}, \
  and\ \bibinfo {author} {\bibfnamefont {J.}~\bibnamefont {Zhou}},\ }\href
  {\doibase 10.1103/PhysRevD.77.125010} {\bibfield  {journal} {\bibinfo
  {journal} {Phys. Rev. D}\ }\textbf {\bibinfo {volume} {77}},\ \bibinfo
  {pages} {125010} (\bibinfo {year} {2008})}\BibitemShut {NoStop}%
\bibitem [{\citenamefont {Sch\"afer}\ and\ \citenamefont
  {Zhou}(2013)}]{Schafer:2013mza}%
  \BibitemOpen
  \bibfield  {author} {\bibinfo {author} {\bibfnamefont {A.}~\bibnamefont
  {Sch\"afer}}\ and\ \bibinfo {author} {\bibfnamefont {J.}~\bibnamefont
  {Zhou}},\ }\href {\doibase 10.1103/PhysRevD.88.074012} {\bibfield  {journal}
  {\bibinfo  {journal} {Phys. Rev. D}\ }\textbf {\bibinfo {volume} {88}},\
  \bibinfo {pages} {074012} (\bibinfo {year} {2013})}\BibitemShut {NoStop}%
\bibitem [{\citenamefont {Boer}\ \emph {et~al.}(2015)\citenamefont {Boer},
  \citenamefont {Buffing},\ and\ \citenamefont {Mulders}}]{Boer:2015kxa}%
  \BibitemOpen
  \bibfield  {author} {\bibinfo {author} {\bibfnamefont {D.}~\bibnamefont
  {Boer}}, \bibinfo {author} {\bibfnamefont {M.}~\bibnamefont {Buffing}}, \
  and\ \bibinfo {author} {\bibfnamefont {P.}~\bibnamefont {Mulders}},\ }\href
  {\doibase 10.1007/JHEP08(2015)053} {\bibfield  {journal} {\bibinfo  {journal}
  {JHEP}\ }\textbf {\bibinfo {volume} {08}},\ \bibinfo {pages} {053} (\bibinfo
  {year} {2015})}\BibitemShut {NoStop}%
\bibitem [{\citenamefont {Zhang}\ \emph {et~al.}(2019)\citenamefont {Zhang},
  \citenamefont {Qin},\ and\ \citenamefont {Wang}}]{Zhang:2019toi}%
  \BibitemOpen
  \bibfield  {author} {\bibinfo {author} {\bibfnamefont {Y.-Y.}\ \bibnamefont
  {Zhang}}, \bibinfo {author} {\bibfnamefont {G.-Y.}\ \bibnamefont {Qin}}, \
  and\ \bibinfo {author} {\bibfnamefont {X.-N.}\ \bibnamefont {Wang}},\ }\href
  {\doibase 10.1103/PhysRevD.100.074031} {\bibfield  {journal} {\bibinfo
  {journal} {Phys. Rev. D}\ }\textbf {\bibinfo {volume} {100}},\ \bibinfo
  {pages} {074031} (\bibinfo {year} {2019})}\BibitemShut {NoStop}%
\bibitem [{\citenamefont {Alrashed}\ \emph {et~al.}(2021)\citenamefont
  {Alrashed}, \citenamefont {Anderle}, \citenamefont {Kang}, \citenamefont
  {Terry},\ and\ \citenamefont {Xing}}]{Alrashed:2021csd}%
  \BibitemOpen
  \bibfield  {author} {\bibinfo {author} {\bibfnamefont {M.}~\bibnamefont
  {Alrashed}}, \bibinfo {author} {\bibfnamefont {D.}~\bibnamefont {Anderle}},
  \bibinfo {author} {\bibfnamefont {Z.-B.}\ \bibnamefont {Kang}}, \bibinfo
  {author} {\bibfnamefont {J.}~\bibnamefont {Terry}}, \ and\ \bibinfo {author}
  {\bibfnamefont {H.}~\bibnamefont {Xing}},\ }\href@noop {} {\  (\bibinfo
  {year} {2021})},\ \Eprint {http://arxiv.org/abs/2107.12401} {arXiv:2107.12401
  [hep-ph]} \BibitemShut {NoStop}%
\bibitem [{\citenamefont {Connors}\ \emph {et~al.}(2018)\citenamefont
  {Connors}, \citenamefont {Nattrass}, \citenamefont {Reed},\ and\
  \citenamefont {Salur}}]{Connors:2017ptx}%
  \BibitemOpen
  \bibfield  {author} {\bibinfo {author} {\bibfnamefont {M.}~\bibnamefont
  {Connors}}, \bibinfo {author} {\bibfnamefont {C.}~\bibnamefont {Nattrass}},
  \bibinfo {author} {\bibfnamefont {R.}~\bibnamefont {Reed}}, \ and\ \bibinfo
  {author} {\bibfnamefont {S.}~\bibnamefont {Salur}},\ }\href {\doibase
  10.1103/RevModPhys.90.025005} {\bibfield  {journal} {\bibinfo  {journal}
  {Rev. Mod. Phys.}\ }\textbf {\bibinfo {volume} {90}},\ \bibinfo {pages}
  {025005} (\bibinfo {year} {2018})}\BibitemShut {NoStop}%
\bibitem [{\citenamefont {Fialkowski}\ and\ \citenamefont
  {Wit}(2007)}]{Fialkowski:2007pt}%
  \BibitemOpen
  \bibfield  {author} {\bibinfo {author} {\bibfnamefont {K.}~\bibnamefont
  {Fialkowski}}\ and\ \bibinfo {author} {\bibfnamefont {R.}~\bibnamefont
  {Wit}},\ }\href {\doibase 10.1140/epja/i2007-10361-2} {\bibfield  {journal}
  {\bibinfo  {journal} {Eur. Phys. J.}\ }\textbf {\bibinfo {volume} {A32}},\
  \bibinfo {pages} {213} (\bibinfo {year} {2007})}\BibitemShut {NoStop}%
%%CITATION = HEP-PH/0702058;%%
\bibitem [{\citenamefont {Airapetian}\ \emph {et~al.}(2006)\citenamefont
  {Airapetian} \emph {et~al.}}]{Airapetian:2005yh}%
  \BibitemOpen
  \bibfield  {author} {\bibinfo {author} {\bibfnamefont {A.}~\bibnamefont
  {Airapetian}} \emph {et~al.} (\bibinfo {collaboration} {\textit{HERMES
  Collaboration}}),\ }\href {\doibase 10.1103/PhysRevLett.96.162301} {\bibfield
   {journal} {\bibinfo  {journal} {Phys. Rev. Lett.}\ }\textbf {\bibinfo
  {volume} {96}},\ \bibinfo {pages} {162301} (\bibinfo {year}
  {2006})}\BibitemShut {NoStop}%
%%CITATION = HEP-EX/0510030;%%
\bibitem [{\citenamefont {Agababyan}\ \emph {et~al.}(2011)\citenamefont
  {Agababyan}, \citenamefont {Ammosov}, \citenamefont {Atayan}, \citenamefont
  {Grigoryan}, \citenamefont {Grigoryan}, \citenamefont {Gulkanyan},
  \citenamefont {Ivanilov}, \citenamefont {Karamyan},\ and\ \citenamefont
  {Korotkov}}]{Agababyan:2006ne}%
  \BibitemOpen
  \bibfield  {author} {\bibinfo {author} {\bibfnamefont {N.}~\bibnamefont
  {Agababyan}}, \bibinfo {author} {\bibfnamefont {V.}~\bibnamefont {Ammosov}},
  \bibinfo {author} {\bibfnamefont {M.}~\bibnamefont {Atayan}}, \bibinfo
  {author} {\bibfnamefont {L.}~\bibnamefont {Grigoryan}}, \bibinfo {author}
  {\bibfnamefont {N.}~\bibnamefont {Grigoryan}}, \bibinfo {author}
  {\bibfnamefont {H.}~\bibnamefont {Gulkanyan}}, \bibinfo {author}
  {\bibfnamefont {A.}~\bibnamefont {Ivanilov}}, \bibinfo {author}
  {\bibfnamefont {Z.}~\bibnamefont {Karamyan}}, \ and\ \bibinfo {author}
  {\bibfnamefont {V.}~\bibnamefont {Korotkov}},\ }\href {\doibase
  10.1134/S1063778811020050} {\bibfield  {journal} {\bibinfo  {journal} {Phys.
  Atom. Nucl.}\ }\textbf {\bibinfo {volume} {74}},\ \bibinfo {pages} {246}
  (\bibinfo {year} {2011})}\BibitemShut {NoStop}%
\bibitem [{\citenamefont {Hakobyan}(2008)}]{Hakobyan:2008kua}%
  \BibitemOpen
  \bibfield  {author} {\bibinfo {author} {\bibfnamefont {H.}~\bibnamefont
  {Hakobyan}},\ }\emph {\bibinfo {title} {{Observation of Quark Propagation
  Pattern in Nuclear Medium}}},\ \href
  {http://www.jlab.org/Hall-B/general/thesis/Hakobyan_thesis.pdf} {Ph.D.
  thesis},\ \bibinfo  {school} {Yerevan State U.} (\bibinfo {year}
  {2008})\BibitemShut {NoStop}%
%%CITATION = INSPIRE-1288986;%%
\bibitem [{\citenamefont {Mecking}\ \emph {et~al.}(2003)\citenamefont {Mecking}
  \emph {et~al.}}]{Mecking:2003zu}%
  \BibitemOpen
  \bibfield  {author} {\bibinfo {author} {\bibfnamefont {B.~A.}\ \bibnamefont
  {Mecking}} \emph {et~al.} (\bibinfo {collaboration} {\textit{CLAS
  Collaboration}}),\ }\href {\doibase 10.1016/S0168-9002(03)01001-5} {\bibfield
   {journal} {\bibinfo  {journal} {Nucl. Instrum. Meth.}\ }\textbf {\bibinfo
  {volume} {A503}},\ \bibinfo {pages} {513} (\bibinfo {year}
  {2003})}\BibitemShut {NoStop}%
%%CITATION = NUIMA,A503,513;%%
\bibitem [{\citenamefont {El~Fassi}\ \emph {et~al.}(2012)\citenamefont
  {El~Fassi} \emph {et~al.}}]{ElFassi:2012nr}%
  \BibitemOpen
  \bibfield  {author} {\bibinfo {author} {\bibfnamefont {L.}~\bibnamefont
  {El~Fassi}} \emph {et~al.} (\bibinfo {collaboration} {\textit{CLAS
  Collaboration}}),\ }\href {\doibase 10.1016/j.physletb.2012.05.019}
  {\bibfield  {journal} {\bibinfo  {journal} {Phys. Lett.}\ }\textbf {\bibinfo
  {volume} {B712}},\ \bibinfo {pages} {326} (\bibinfo {year}
  {2012})}\BibitemShut {NoStop}%
%%CITATION = ARXIV:1201.2735;%%
\bibitem [{\citenamefont {Hakobyan}\ \emph {et~al.}(2008)\citenamefont
  {Hakobyan} \emph {et~al.}}]{Hakobyan:2008zz}%
  \BibitemOpen
  \bibfield  {author} {\bibinfo {author} {\bibfnamefont {H.}~\bibnamefont
  {Hakobyan}} \emph {et~al.},\ }\href {\doibase 10.1016/j.nima.2008.04.055}
  {\bibfield  {journal} {\bibinfo  {journal} {Nucl. Instrum. Meth.}\ }\textbf
  {\bibinfo {volume} {A592}},\ \bibinfo {pages} {218} (\bibinfo {year}
  {2008})}\BibitemShut {NoStop}%
%%CITATION = NUIMA,A592,218;%%
\bibitem [{sup()}]{supplementary}%
  \BibitemOpen
  \href@noop {} {}\bibinfo {note} {See Supplemental Material at [insert URL
  here] for tables of the conditional modification factor, statistical and
  systematic uncertainties, and the bin edges.}\BibitemShut {Stop}%
\bibitem [{\citenamefont {Wolin}(1995)}]{GSIM}%
  \BibitemOpen
  \bibfield  {author} {\bibinfo {author} {\bibfnamefont {E.}~\bibnamefont
  {Wolin}},\ }\href@noop {} {\enquote {\bibinfo {title} {{GSIM User’s Guide
  Version CERN 1.0}},}\ } (\bibinfo {year} {1995})\BibitemShut {NoStop}%
%%CITATION = CERN-W5013;%%
\bibitem [{\citenamefont {Brun}\ \emph {et~al.}(1994)\citenamefont {Brun},
  \citenamefont {Bruyant}, \citenamefont {Carminati}, \citenamefont {Giani},
  \citenamefont {Maire}, \citenamefont {McPherson}, \citenamefont {Patrick},\
  and\ \citenamefont {Urban}}]{Brun:1994aa}%
  \BibitemOpen
  \bibfield  {author} {\bibinfo {author} {\bibfnamefont {R.}~\bibnamefont
  {Brun}}, \bibinfo {author} {\bibfnamefont {F.}~\bibnamefont {Bruyant}},
  \bibinfo {author} {\bibfnamefont {F.}~\bibnamefont {Carminati}}, \bibinfo
  {author} {\bibfnamefont {S.}~\bibnamefont {Giani}}, \bibinfo {author}
  {\bibfnamefont {M.}~\bibnamefont {Maire}}, \bibinfo {author} {\bibfnamefont
  {A.}~\bibnamefont {McPherson}}, \bibinfo {author} {\bibfnamefont
  {G.}~\bibnamefont {Patrick}}, \ and\ \bibinfo {author} {\bibfnamefont
  {L.}~\bibnamefont {Urban}},\ }\href {\doibase 10.17181/CERN.MUHF.DMJ1} {\
  (\bibinfo {year} {1994}),\ 10.17181/CERN.MUHF.DMJ1}\BibitemShut {NoStop}%
%%CITATION = CERN-W5013;%%
\bibitem [{\citenamefont {Buss}\ \emph {et~al.}(2012)\citenamefont {Buss},
  \citenamefont {Gaitanos}, \citenamefont {Gallmeister}, \citenamefont {van
  Hees}, \citenamefont {Kaskulov}, \citenamefont {Lalakulich}, \citenamefont
  {Larionov}, \citenamefont {Leitner}, \citenamefont {Weil},\ and\
  \citenamefont {Mosel}}]{Buss:2011mx}%
  \BibitemOpen
  \bibfield  {author} {\bibinfo {author} {\bibfnamefont {O.}~\bibnamefont
  {Buss}}, \bibinfo {author} {\bibfnamefont {T.}~\bibnamefont {Gaitanos}},
  \bibinfo {author} {\bibfnamefont {K.}~\bibnamefont {Gallmeister}}, \bibinfo
  {author} {\bibfnamefont {H.}~\bibnamefont {van Hees}}, \bibinfo {author}
  {\bibfnamefont {M.}~\bibnamefont {Kaskulov}}, \bibinfo {author}
  {\bibfnamefont {O.}~\bibnamefont {Lalakulich}}, \bibinfo {author}
  {\bibfnamefont {A.~B.}\ \bibnamefont {Larionov}}, \bibinfo {author}
  {\bibfnamefont {T.}~\bibnamefont {Leitner}}, \bibinfo {author} {\bibfnamefont
  {J.}~\bibnamefont {Weil}}, \ and\ \bibinfo {author} {\bibfnamefont
  {U.}~\bibnamefont {Mosel}},\ }\href {\doibase 10.1016/j.physrep.2011.12.001}
  {\bibfield  {journal} {\bibinfo  {journal} {Phys. Rept.}\ }\textbf {\bibinfo
  {volume} {512}},\ \bibinfo {pages} {1} (\bibinfo {year} {2012})}\BibitemShut
  {NoStop}%
%%CITATION = ARXIV:1106.1344;%%
\bibitem [{\citenamefont {Airapetian}\ \emph
  {et~al.}(2010{\natexlab{b}})\citenamefont {Airapetian} \emph
  {et~al.}}]{HERMES:2009uge}%
  \BibitemOpen
  \bibfield  {author} {\bibinfo {author} {\bibfnamefont {A.}~\bibnamefont
  {Airapetian}} \emph {et~al.} (\bibinfo {collaboration} {\textit{HERMES
  Collaboration}}),\ }\href {\doibase 10.1016/j.physletb.2010.01.020}
  {\bibfield  {journal} {\bibinfo  {journal} {Phys. Lett. B}\ }\textbf
  {\bibinfo {volume} {684}},\ \bibinfo {pages} {114} (\bibinfo {year}
  {2010}{\natexlab{b}})},\ \Eprint {http://arxiv.org/abs/0906.2478}
  {arXiv:0906.2478 [hep-ex]} \BibitemShut {NoStop}%
\bibitem [{\citenamefont {Majumder}\ and\ \citenamefont
  {Wang}(2008)}]{Majumder:2008jy}%
  \BibitemOpen
  \bibfield  {author} {\bibinfo {author} {\bibfnamefont {A.}~\bibnamefont
  {Majumder}}\ and\ \bibinfo {author} {\bibfnamefont {X.-N.}\ \bibnamefont
  {Wang}},\ }\href@noop {} {\  (\bibinfo {year} {2008})},\ \Eprint
  {http://arxiv.org/abs/0806.2653} {arXiv:0806.2653 [nucl-th]} \BibitemShut
  {NoStop}%
\bibitem [{\citenamefont {Majumder}(2012)}]{Majumder:2009ge}%
  \BibitemOpen
  \bibfield  {author} {\bibinfo {author} {\bibfnamefont {A.}~\bibnamefont
  {Majumder}},\ }\href {\doibase 10.1103/PhysRevD.85.014023} {\bibfield
  {journal} {\bibinfo  {journal} {Phys. Rev. D}\ }\textbf {\bibinfo {volume}
  {85}},\ \bibinfo {pages} {014023} (\bibinfo {year} {2012})},\ \Eprint
  {http://arxiv.org/abs/0912.2987} {arXiv:0912.2987 [nucl-th]} \BibitemShut
  {NoStop}%
\bibitem [{\citenamefont {Gallmeister}\ and\ \citenamefont
  {Mosel}(2008)}]{GALLMEISTER200868}%
  \BibitemOpen
  \bibfield  {author} {\bibinfo {author} {\bibfnamefont {K.}~\bibnamefont
  {Gallmeister}}\ and\ \bibinfo {author} {\bibfnamefont {U.}~\bibnamefont
  {Mosel}},\ }\href {\doibase https://doi.org/10.1016/j.nuclphysa.2007.12.009}
  {\bibfield  {journal} {\bibinfo  {journal} {Nuclear Physics A}\ }\textbf
  {\bibinfo {volume} {801}},\ \bibinfo {pages} {68} (\bibinfo {year}
  {2008})}\BibitemShut {NoStop}%
\bibitem [{\citenamefont {Larionov}\ and\ \citenamefont
  {Strikman}(2020)}]{PhysRevC.101.014617}%
  \BibitemOpen
  \bibfield  {author} {\bibinfo {author} {\bibfnamefont {A.~B.}\ \bibnamefont
  {Larionov}}\ and\ \bibinfo {author} {\bibfnamefont {M.}~\bibnamefont
  {Strikman}},\ }\href {\doibase 10.1103/PhysRevC.101.014617} {\bibfield
  {journal} {\bibinfo  {journal} {Phys. Rev. C}\ }\textbf {\bibinfo {volume}
  {101}},\ \bibinfo {pages} {014617} (\bibinfo {year} {2020})}\BibitemShut
  {NoStop}%
\bibitem [{\citenamefont {Burkert}(2018)}]{Burkert:2018nvj}%
  \BibitemOpen
  \bibfield  {author} {\bibinfo {author} {\bibfnamefont {V.~D.}\ \bibnamefont
  {Burkert}},\ }\href {\doibase 10.1146/annurev-nucl-101917-021129} {\bibfield
  {journal} {\bibinfo  {journal} {Ann. Rev. Nucl. Part. Sci.}\ }\textbf
  {\bibinfo {volume} {68}},\ \bibinfo {pages} {405} (\bibinfo {year}
  {2018})}\BibitemShut {NoStop}%
\bibitem [{\citenamefont {Burkert}\ \emph {et~al.}(2020)\citenamefont {Burkert}
  \emph {et~al.}}]{Burkert:2020akg}%
  \BibitemOpen
  \bibfield  {author} {\bibinfo {author} {\bibfnamefont {V.~D.}\ \bibnamefont
  {Burkert}} \emph {et~al.},\ }\href {\doibase 10.1016/j.nima.2020.163419}
  {\bibfield  {journal} {\bibinfo  {journal} {Nucl. Instrum. Meth. A}\ }\textbf
  {\bibinfo {volume} {959}},\ \bibinfo {pages} {163419} (\bibinfo {year}
  {2020})}\BibitemShut {NoStop}%
\bibitem [{\citenamefont {Arrington}\ \emph {et~al.}(2021)\citenamefont
  {Arrington} \emph {et~al.}}]{Arrington:2021alx}%
  \BibitemOpen
  \bibfield  {author} {\bibinfo {author} {\bibfnamefont {J.}~\bibnamefont
  {Arrington}} \emph {et~al.},\ }\href@noop {} {\  (\bibinfo {year} {2021})},\
  \Eprint {http://arxiv.org/abs/2112.00060} {arXiv:2112.00060 [nucl-ex]}
  \BibitemShut {NoStop}%
\bibitem [{\citenamefont {Abdul~Khalek}\ \emph {et~al.}(2021)\citenamefont
  {Abdul~Khalek} \emph {et~al.}}]{AbdulKhalek:2021gbh}%
  \BibitemOpen
  \bibfield  {author} {\bibinfo {author} {\bibfnamefont {R.}~\bibnamefont
  {Abdul~Khalek}} \emph {et~al.},\ }\href@noop {} {\  (\bibinfo {year}
  {2021})},\ \Eprint {http://arxiv.org/abs/2103.05419} {arXiv:2103.05419
  [physics.ins-det]} \BibitemShut {NoStop}%
\bibitem [{\citenamefont {Anderle}\ \emph {et~al.}(2021)\citenamefont {Anderle}
  \emph {et~al.}}]{Anderle:2021wcy}%
  \BibitemOpen
  \bibfield  {author} {\bibinfo {author} {\bibfnamefont {D.~P.}\ \bibnamefont
  {Anderle}} \emph {et~al.},\ }\href {\doibase 10.1007/s11467-021-1062-0}
  {\bibfield  {journal} {\bibinfo  {journal} {Front. Phys. (Beijing)}\ }\textbf
  {\bibinfo {volume} {16}},\ \bibinfo {pages} {64701} (\bibinfo {year}
  {2021})}\BibitemShut {NoStop}%
\end{thebibliography}%



%\begin{thebibliography}{}
%\input{biblio.bbl}
%\end{thebibliography}

%% \pagebreak
%% \appendix*
%% \input{sections/appendix1.tex}

\end{document}